\documentstyle[12pt]{article}
\catcode`@=11
\newcount\@tempcntc
\def\@citex[#1]#2{\if@filesw\immediate\write\@auxout{\string\citation{#2}}\fi
  \@tempcnta\z@\@tempcntb\m@ne\def\@citea{}\@cite{\@for\@citeb:=#2\do
    {\@ifundefined
       {b@\@citeb}{\@citeo\@tempcntb\m@ne\@citea\def\@citea{,}{\bf ?}\@warning
       {Citation `\@citeb' on page \thepage \space undefined}}%
    {\setbox\z@\hbox{\global\@tempcntc0\csname b@\@citeb\endcsname\relax}%
     \ifnum\@tempcntc=\z@ \@citeo\@tempcntb\m@ne
       \@citea\def\@citea{,}\hbox{\csname b@\@citeb\endcsname}%
     \else
      \advance\@tempcntb\@ne
      \ifnum\@tempcntb=\@tempcntc
      \else\advance\@tempcntb\m@ne\@citeo
      \@tempcnta\@tempcntc\@tempcntb\@tempcntc\fi\fi}}\@citeo}{#1}}
\def\@citeo{\ifnum\@tempcnta>\@tempcntb\else\@citea\def\@citea{,}%
  \ifnum\@tempcnta=\@tempcntb\the\@tempcnta\else
   {\advance\@tempcnta\@ne\ifnum\@tempcnta=\@tempcntb \else \def\@citea{--}\fi
    \advance\@tempcnta\m@ne\the\@tempcnta\@citea\the\@tempcntb}\fi\fi}
\catcode`@=12
\def\aa{{\mbox{\scriptsize\sf a}}}
\def\bb{{\mbox{\scriptsize\sf b}}}
\def\cc{{\mbox{\scriptsize\sf c}}}
\def\a{\alpha}      \def\b{\beta}     \def\c{\gamma}      \def\d{\delta}
\def\e{\epsilon}        \def\l{\lambda}     \def\m{\mu}
\def\n{\nu}         \def\o{\omega}             
\def\s{\sigma}            \def\z{\zeta}

\def\Dst{\displaystyle}
\def\Tst{\textstyle}

\def\SSst{\scriptscriptstyle}
\def\fff#1{\mbox{\boldmath$#1$}}

\def\beq{\begin{equation}}            \def\eeq{\end{equation}}
\def\baq{\begin{eqnarray}}            \def\eaq{\end{eqnarray}}
\def\baqn{\begin{eqnarray*}}          \def\eaqn{\end{eqnarray*}}

\def\real{\mbox{$\mbox{I}\!\mbox{R}$}}
\def\complex{\mbox{$\,\,\mbox{\rule{0.06em}{1.4ex}}\!\!\mbox{C}$}}
\def\One{\mbox{$1\!\!\!\;\mbox{l}$}}
\def\glfourr{\mbox{$\mbox{GL}(4,\!\real)$}}
\def\glfourc{\mbox{$\mbox{GL}(4,\!\complex)$}}
\def\sltwoc{\mbox{$\mbox{SL}(2,\!\complex)$}}
\def\glonec{\mbox{$\mbox{GL}(1,\!\complex)$}}
\def\uone{\mbox{$\mbox{U}(1)$}}
\def\cspin{\mbox{$\complex\mbox{Spin}$}}

\def\clplus{\mbox{$\complex L^+$}}
\def\lplus{\mbox{$L^+_\uparrow$}}

\def\lplusm{\mbox{$L^+_\uparrow(M)$}}
\def\clplusm{\mbox{$\complex L^+(M)$}}
\def\cspinm{\mbox{$\complex\mbox{Spin}(M)$}}

\def\cxm{\mbox{$\complex^\times(M)$}}

\def\Con#1#2#3{\Gamma^#1{}_{\! #2 #3}}
\def\con#1#2#3{\Gamma_{#1 #2 #3}}
\def\Chr#1#2#3{\big\{{}^#1{}_{#2 #3}\big\}}
\def\chr#1#2#3{\big\{{}_{#1 #2 #3}\big\}}

\def\PSI{\overline{\psi}} \def\psid{\psi_z}
\def\planck{l_{\SSst 0}}  \def\PSId{\PSI_z}
\def\sumd{\sum^{}_{z}}
\def\currentd{j_z}

\begin{document}
\thispagestyle{empty}

\begin{flushright}
                           MZ--TH/94--16\\
                           May 1994\\
\end{flushright}

\begin{center}
{\bf\Large         Geometric Interpretation of        }\\[3mm]
{\bf\Large Electromagnetism in a Gravitational Theory }\\[3mm]
{\bf\Large           with Space--Time Torsion         }\\[13mm]

{\large                    Kenichi Horie
\footnote{e--mail: horie@vipmzw.physik.uni-mainz.de}  }\\[3mm]

Institut f\"ur Physik, Johannes Gutenberg--Universit\"at Mainz\\
              D--55099 Mainz, Germany                         \\[13mm]
\end{center}

{\bf\large Abstract}\\[1mm]

A complete geometric unification of gravity and electromagnetism is proposed
by considering two aspects of torsion: its relation to spin
established in Einstein--Cartan theory and the possible interpretation of the
torsion trace as the electromagnetic potential. Starting with a Lagrangian
built of Dirac spinors, orthonormal tetrads, and a complex rather than a real
linear connection we define an extended spinor derivative by which we obtain
not only a very natural unification, but can also fully clarify the nontrivial
underlying fibre bundle structure. Thereby a new type of contact interaction
between spinors emerges, which differs from the usual one in Einstein--Cartan
theory. The splitting of the linear connection into a metric and an
electromagnetic part together with a characteristic length scale in the
theory strongly suggest that gravity and electromagnetism have the same
geometrical origin.\\[1mm]


\newpage

%%%%%%%%%%%%%%%%%%%%%%%%%%%%%%%%%%%%%%%%%%%%%%%%%%%%%%%%%%%%%%%%%%%%%%%%%%%%
%
                          \section{Introduction}
%
%%%%%%%%%%%%%%%%%%%%%%%%%%%%%%%%%%%%%%%%%%%%%%%%%%%%%%%%%%%%%%%%%%%%%%%%%%%%

In general relativity the metric $g_{\m\n}$ completely determines the
linear connection $\Con\a\m\b$, which becomes simply the
(symmetric) Levi--Civita connection
\beq
  \Con\a\m\b = \Chr\a\m\b := {\Tst\frac{1}{2}}g^{\a\e}
  \left(\partial_\m g_{\e\b}+\partial_\b g_{\e\m}-\partial_\e g_{\m\b}
  \right)\;\;.
\eeq
The space--time geometry is influenced only by mass--energy, which causes
curvature via the Einstein equation, and remains unaffected by spin.

In Einstein--Cartan theory (see Ref.\ \cite{1} and references therein) the
connection is only required to be metric, $\nabla_{\!\a}g_{\m\n}=0$, and is
allowed to have nonvanishing torsion $T^\a{}_{\!\m\b}=\Con\a\m\b-\Con\a\b\m$,
contrary to the torsionless Levi--Civita connection. The structure of the
connection now becomes \cite{1}
\beq
  \Con\a\m\b=\Chr\a\m\b + {\Tst\frac{1}{2}}
             (T_\m{}^\a{}_{\!\b}+T_\b{}^\a{}_{\!\m}+T^\a{}_{\!\m\b}) \;\;.
\eeq
This generalization enables the space--time geometry to respond not only to
mass but also to spin, where spinning matter produces torsion. For Dirac
particles the torsion is totally antisymmetric in its indices and creates
a cubic self-interaction term in the spinor equation \cite{2}
\beq
    i\c^\m\nabla^\ast_{\!\m}\psi
   -\frac{mc}{\hbar}\psi
   +{\Tst\frac{3}{8}}\planck^2 (\PSI\c^5\c^\d\psi)\c^5\c_\d\psi
 = 0
\eeq
where $\nabla^\ast_{\!\m}$ is the covariant spinor derivative with respect
to the Levi--Civita connection, see (3.12), and $\planck$ is
the Planck length. The observed ``contact interaction'' in (1.3) also
contributes to the energy--momentum equation \cite{2}.

Besides this well-known aspect of torsion another physical role for it
has been suggested in several works on the unification of gravitation and
electromagnetism. The idea of such a geometrical unification is to omit any
restrictions on the real linear connection $\Con\a\m\b$ and to incorporate
the electromagnetic phenomena into this extended space--time geometry. More
precisely, the electromagnetic vector potential $A_\m$ is identified with
the torsion trace $T_\m=T^\a{}_{\!\m\a}$. In the so called nonsymmetric
unified field theory, Einstein \cite{3} has considered a general linear
connection, but his aim was to incorporate electromagnetism into the metric.
He introduced a nonsymmetric metric ${\tilde{g}}_{\m\n}
(\neq {\tilde{g}}_{\n\m})$ and identified its antisymmetric part with the
dual of the electromagnetic field strength. His theory was unsuccessful
because it could not account for the equation of motion \cite{4,5}. To remedy
this and various other shortcomings of Einstein's theory several authors have
suggested to make the above mentioned identification $T_\m\sim A_\m$ in an
ad hoc manner \cite{6,7}, still using a nonsymmetric metric. In subsequent
works of McKellar \cite{8} and Jakubiec and Kijowski \cite{10} this is
achieved without ad hoc assumptions, using the usual symmetric metric.
McKellar starts with the metric $g_{\m\n}$ and a general linear connection
$\Con\a\m\b$ and obtains
\beq
  \Con\a\m\b = \Chr\a\m\b +{\Tst\frac{1}{3}}\d^\a{}_{\!\b}\cdot T_\m
\eeq
as solution of the field equation with the usual
convention $\nabla_{\!\m}X^\a=\partial_\m X^\a+\Con\a\m\b
X^\b$ for the covariant derivative of a vector field $X^\a$. His field
equations taken together resemble the source-free Einstein--Maxwell
equations, provided that $T_\m \sim A_\m$ holds. In Ref.\ \cite{9} Ferraris
and Kijowski start not with a metric but
with $\Con\a\m\b$ alone and arrive at (1.4) but,
contrary to McKellar, they regard $\Con\a\m\a$, which is not a vector, as
the electromagnetic potential and deduce a theory of electromagnetism
differing from the Maxwell theory, whereas in Ref.\ \cite{10} Jakubiec and
Kijowski return to the identity $T_\m \sim A_\m$ and include Dirac spinors in
the unification. Unfortunately the employed spinor derivative requires two
connections $\Con\a\m\b$ and $\Chr\a\m\b$ from the very beginning, and
furthermore, from the general linear connection $\Con\a\m\b$ only its trace
$\Con\a\m\a$ appears in this derivative. Owing to this insufficient
coupling of the linear connection to spinorial matter torsion does not
couple to spin and so its important physical role established in
Einstein--Cartan theory is missing. Like McKellar, the authors of Ref.\
\cite{10} do not clarify the fibre bundle structure of their unification;
for example, $T_\m$ in (1.4) cannot be gauged with \uone, as there is no
\uone--bundle constructed in the theory. It is thus only a vector but not
a potential.

{}From the discussions of general relativity, Einstein--Cartan theory, and
the unified field theories we conclude that a more general linear connection
would enable the space--time geometry to incorporate further physical
phenomena in addition to gravitation. However, although the \glfourr--%
connection of Refs.\ \cite{8,9,10} is more general than the metric
connection of Einstein--Cartan theory the spin-torsion coupling was missing
either because matter was not considered \cite{8,9} or because the spinor
derivative was somewhat inappropriate \cite{10}.

In this work our aim is to obtain a new unification of gravity and
electromagnetism
including the spin-torsion coupling, thus accounting for both aspects of
torsion mentioned before. To achieve this we further expand the space-time
geometry and allow for complex linear connections. Let us explain why this
complex extension is necessary. Obviously, we must introduce a new spinor
derivative containing a spin-torsion coupling. As a consequence,
we may imagine that matter
will be coupled to the connection more tightly and can therefore twist the
space--time geometry so strongly that even complex degrees of freedom are
excited. Another reason for the complex extension is the fact that in Refs.\
\cite{8,10} no \uone--bundle structure could be constructed for the torsion
trace $T_\m$ of a real connection.

Using a complex linear connection and an extended spinor derivative we arrive
at a new unification of gravity and electromagnetism which fully clarifies
the fibre bundle structure, especially the \uone--bundle. All field
equations follow from the variational principle. Due to the special spinor
derivative both aspects of torsion in Eqs.\ (1.3) and (1.4) are slightly
altered. The spinor--spinor contact interaction is now found to occur only
between distinct particles, thus excluding a self-interaction like in
Eq.\ (1.3).

In Section 2 we establish notation and introduce the Lagrangian density.
In Sections 3 and 4 the field equations are derived using the variational
method and their physical content is discussed. Here we
slightly expand the theory to include three types of charged particles
and also consider many particle systems to observe the new type of
spinor--spinor contact interaction. In Section 5 a short summary is given.
The fibre bundle geometry is discussed in Appendix A.

\setcounter{equation}{0}

%%%%%%%%%%%%%%%%%%%%%%%%%%%%%%%%%%%%%%%%%%%%%%%%%%%%%%%%%%%%%%%%%%%%%%%%%%
%
                 \section{Lagrangian density}
%
%%%%%%%%%%%%%%%%%%%%%%%%%%%%%%%%%%%%%%%%%%%%%%%%%%%%%%%%%%%%%%%%%%%%%%%%%%

The theory rests on the variational principle and employs a Lagrangian
density built of orthonormal tetrads, a complex linear connection, and
Dirac spinors. To introduce these field variables let us consider a real
4--dimensional space--time manifold $M$. We assume that $M$ is endowed with
a pseudo-riemannian metric $g_{\m\n}$ and a space-- and time--orientation which
follows from the
reduction of the frame bundle $F(M)$ to a special Lorentz bundle \lplusm;
this is a principal bundle consisting of orthonormal tangent bases
such that the structure group is given by the special orthochronous
Lorentz group \lplus\ with Lie algebra \fff l \\
\parbox{1cm}{ }
\hfill
\parbox{11cm}{
  \baqn
    \lplus
  &:=&
    \left\{ \Lambda \in \mbox{Mat}(4,\!\real)| \Lambda^T\eta\Lambda=\eta\,,
    \:\det\Lambda =1\,,\: \Lambda^0{}_{\!0}\ge1 \right\} \;\;;
  \\
    \fff l
  &=&
    \left\{ \Lambda \in \mbox{Mat}(4,\!\real)| \Lambda^T\eta+\eta\Lambda=0
    \right\} \;\;,
  \eaqn }
\hfill
\parbox{1cm}{\baq\eaq}\\
where $\eta=(\eta_{\aa\bb})=(\eta^{\aa\bb})=\mbox{diag}(1,-1,-1,-1)$. A tetrad
$\s=(e_\aa{}^{\!\m}\partial_\m)$ is a local cross section in \lplusm,
where latin indices,
running from 0 to 3, are anholonomic and will be lowered and raised with
$\eta_{\aa\bb}$ and $\eta^{\aa\bb}$, respectively. Greek indices run also
from 0 to 3 and refer to local coordinates. They are lowered and raised with
$g_{\m\n}$ and $g^{\m\n}$, the latter being the inverse of $g_{\m\n}$. Let
$(e^\aa{}_{\!\m}dx^\m)$ denote the reciprocal tetrad satisfying
$e_\aa{}^{\!\m}e^\aa{}_{\!\n}=\d^\m{}_{\!\n}$ and
$e_\aa{}^{\!\m}e^\bb{}_{\!\m}=\d^\bb{}_{\!\aa}$. We then have the following
relations
\beq
  g^{\m\n}=e_\aa{}^{\!\m}e^{\aa\n}\;,\quad
  g_{\m\n}=e_{\aa\m}e^\aa{}_{\!\n}\;,\quad
  e:=\det(e^\aa{}_{\!\m})=\sqrt{-\det(g_{\m\n})}\;\;.
\eeq
The components $e_\aa{}^{\!\m}$ and $e^\aa{}_{\!\m}$ will be used to convert
coordinate indices to anholonomic ones and vice versa.

The complex frame bundle $F_c(M)$ is a \glfourc--principal bundle
consisting of all complex tangent bases of $\complex\otimes TM$.
In particular a tetrad $\s$ is a cross section in $F_c(M)$. Therefore
a \glfourc--connection $\o$ on $F_c(M)$
can be pulled back to $M$ via $\s$, yielding a $\fff{gl}(4,
\!\complex)$--valued 1-form $(\s^\ast\o)^\aa{}_{\!\bb}=:\Con\aa\m\bb\,dx^\m$,
which we call a complex linear connection. Its coordinate components
$\Con\a\m\b =e_\aa{}^{\!\a}\,e^\bb{}_{\!\b}\Con\aa\m\bb
+e_\cc{}^{\!\a}\partial_\m e^\cc{}_{\!\b}$ transform in the well-known
inhomogeneous way under coordinate
changes. The curvature tensor, Ricci tensor, curvature scalar, and the
curvature trace are defined as follows\\
\parbox{13cm}{
\baqn
    R^\aa{}_{\!\bb\m\n}
  &=&
    \partial_\m\Con\aa\n\bb+\Con\aa\m\cc\Con\cc\n\bb
   -\partial_\n\Con\aa\m\bb-\Con\aa\n\cc\Con\cc\m\bb \;\;;
  \\
    R_{\b\n}
  &=&
    R^\aa{}_{\!\bb\m\n}\cdot e_\aa{}^{\!\m}\,e^\bb{}_{\!\b} \;\;;
  \\
    R
  &=&
    R^\aa{}_{\!\bb\m\n}\cdot e_\aa{}^{\!\m}\,e^{\bb\n} \;\;;
  \\
    Y_{\m\n}
  &=&
    R^\aa{}_{\!\aa\m\n}=\partial_\m\Con\aa\n\aa-\partial_\n\Con\aa\m\aa \;\;.
\eaqn}
\hfill
\parbox{1cm}{\baq\eaq}
Note that $\Con\aa\n\aa$ is a vector, contrary to $\Con\a\n\a$. This vector
vanishes for metric connections on \lplusm\ because
of the Lie algebra condition $\con\aa\m\bb+\con\bb\m\aa=0$.

Every metric connection defines a covariant differentiation of a Dirac
spinor \cite{2}
\beq
  \nabla_{\!\m}\psi =
  \partial_\m\psi - {\Tst\frac{1}{4}}\con\aa\m\bb\c^\bb\c^\aa\psi \;\;,
\eeq
where the $\c$--matrices satisfy
$\c^\aa\c^\bb+\c^\bb\c^\aa=2\eta^{\aa\bb}\One$, see
e.\ g.\ Ref.\ \cite{2}. We extend (2.4) to the case where $\Con\aa\m\bb$ is a
complex linear connection. At this stage (2.4) is rather a formal
definition as it is only \lplus--covariant but not with respect to \glfourc.
The full geometrical meaning of (2.4) is expounded in Appendix A,
where we also clarify the nontrivial bundle geometry of our unification
scheme.

Introducing $\PSI:=\psi^\dagger\c^0$, $\c^\m:=\c^\aa\,e_\aa{}^{\!\m}$, the mass
of the spinor particle $m$, $k=8\pi G / c^4$, and a length scale $l$
we write down the following Lagrangian density
\baq
    {\cal L}
  &=& {\cal L}_m + {\cal L}_G + {\cal L}_Y
  \nonumber\\
  &=:&
    e\cdot \hbar c
    \left[ i\PSI\c^\m\nabla_{\!\m}\psi - \frac{mc}{\hbar}\PSI\psi \right]
   -\frac{e}{2k}R + \frac{e}{4k}l^2 Y_{\m\n}Y^{\m\n} \;\;.
\eaq
Although it is complex valued we do not make it real, because this would
restrict the contributions of the full complex connection. For an
interesting example of complex Lagrangian theory see e.\ g.\  Ref.\
\cite{11}. Apart from being complex the three parts ${\cal L}_m$, ${\cal L}
_G$, and ${\cal L}_Y$ resemble more or less the usual Lagrangian densities
of spinorial matter, gravity, and the electromagnetic field, respectively.
Whereas expressions similar to ${\cal L}_G$ and ${\cal L}_Y$ for a real
connection were already used in Refs.\ \cite{8,9,10}, the matter Lagrangian
${\cal L}_m$ including the extended spinor derivative (2.4) is new and
plays a key role in our unification. Note that the partial derivatives
$\partial_\m$ and the connection have the dimension of inverse length.
Therefore, from purely dimensional arguments, we must introduce
a squared length $l^2$ in ${\cal L}_Y$. To compare
$l$ with the Planck length $\planck:=\sqrt{\hbar ck}$ we rewrite
(2.5) as follows
\beq
  {\cal L} = \frac{e}{k}\cdot\left[
  i\planck^2\,\PSI\c^\m\nabla_{\!\m}\psi - mc^2 k\PSI\psi
 -{\Tst\frac{1}{2}}R +{\Tst\frac{1}{4}}l^2 Y_{\m\n}Y^{\m\n}\right]  \;\;.
\eeq
In the last term we recognize $l^2$ as the self-coupling constant of the
connection, implying that $l$ is an intrinsic length of the space--time
geometry. The first term on the right side of (2.6) reveals $\planck^2$ as
the coupling constant between the connection and matter. But if
we regard Dirac spinors ultimately as geometrical objects, then $\planck$
also is a characteristic unit of the space--time. We therefore expect
$l$ and $\planck$ being of the same magnitude.

\setcounter{equation}{0}

%%%%%%%%%%%%%%%%%%%%%%%%%%%%%%%%%%%%%%%%%%%%%%%%%%%%%%%%%%%%%%%%%%%%%%%%%%
%
                     \section{Field equations}
%
%%%%%%%%%%%%%%%%%%%%%%%%%%%%%%%%%%%%%%%%%%%%%%%%%%%%%%%%%%%%%%%%%%%%%%%%%%

The field equations are obtained by varying $\cal L$ with respect to {\bf (a)}
$\Con\aa\m\bb$, {\bf (b)} $\psi$ and $\PSI$, and {\bf (c)} $e_\aa{}^{\!\m}$.
The Euler--Lagrange equation for a representative field variable $v$ is
\beq
  0 = \frac{\d {\cal L}}{\d v}
   := \frac{\partial {\cal L}}{\partial v}
     -\partial_\n\frac{\partial {\cal L}}{\partial\partial_\n v} \;\;,
\eeq
where higher order derivatives are absent in $\cal L$. In the following we
give only an outline of the computations. For details see Ref.\ \cite{12}.
\\[3mm]
{\bf (a)}\hspace{2mm} Define the following complex valued third rank tensor
\beq
  \Sigma^\a{}_{\!\m\b} := \Con\a\m\b-\Chr\a\m\b \;.
\eeq
A third rank tensor always admits a ``4-vector decomposition''
\beq
  \Sigma_{\a\b\c}=
   Q_\a g_{\b\c}+S_\b g_{\a\c}+g_{\a\b}U_\c-{\Tst\frac{1}{12}}
   \eta_{\a\b\c\d}V^\d
  +\Upsilon_{\a\b\c} \;\;,
\eeq
where the four vectors $Q_\a$, $S_\b$, $U_\c$, $V_\d$ and the ``tensor rest''
$\Upsilon_{\a\b\c}$ are explicitly defined in Appendix B. This tensor rest
satisfies $\Upsilon^\a{}_{\!\a\c}=\Upsilon^\a{}_{\!\c\a}=
\Upsilon_\c{}^\a{}_{\!\a}=\Upsilon_{[\a\b\c]}=0$. We have denoted the
volume element by $\eta_{\a\b\c\d}:=e\cdot\e_{\a\b\c\d}$, wherein
$\e_{\a\b\c\d}$ is totally antisymmetric and $\e_{0123}=1$. The field
equation for $\Con\aa\m\bb$ follows from (2.5) or (2.6), (3.1) and (3.2)
\baq
  0 &\!=\!&
  \frac{\d {\cal L}}{\d\Con\aa\m\bb}\cdot
  \d^\c{}_{\!\m}\,e^{\aa\a}\,e_\bb{}^{\!\b}
  \cdot\frac{k}{e}  \quad\Leftrightarrow\nonumber
\\
  -{\Tst\frac{1}{4}}i\planck^2\,\PSI\c^\c\c^\b\c^\a\psi &\!=\!&
   {\Tst\frac{1}{2}}
    \left[ \Sigma^{\b\e}{}_{\!\e}g^{\a\c}+\Sigma^\e{}_{\!\e}{}^\a g^{\c\b}
          -\Sigma^{\b\a\c}-\Sigma^{\c\b\a}   \right]
  +l^2 g^{\a\b}\,\nabla^\ast_{\!\n}Y^{\n\c} \;.\quad\,
\eaq
Here $\nabla^\ast_{\!\m}$ is the covariant differentiation with respect to
$\Chr\a\m\b$, see Appendix B.
The bracket $[\ldots]$ contains the contributions of ${\cal L}_G$, wherein
terms belonging to the Levi--Civita connection part in (3.2) and terms
created by the last term of (3.1) cancel out completely. We now use
(3.3) and the definitions of the vector current $j^\a:=\PSI\c^\a\psi$
and the axial current $j^{{\SSst 5}\,\d}:=\PSI\c^5\c^\d\psi$, see Appendix B.
Contracting (3.4) successively with $g_{\b\c}$, $g_{\a\c}$, $g_{\a\b}$ and
$1/6 \cdot \eta_{\c\b\a\d}$ we obtain\\
\parbox{5mm}{ }
\hfill
\parbox{13cm}{
  \[\renewcommand{\arraystretch}{1.7}
    \begin{array}{c@{\qquad}r@{\:=\:}l}
      \langle\; g_{\b\c}:\;\rangle & -i\planck^2\cdot j^\a &
      3Q^\a+6U^\a+l^2\nabla^\ast_{\!\n}Y^{\n\a}
    \\
      \langle\; g_{\a\c}:\;\rangle & \frac{1}{2}i\planck^2\cdot j^\b &
      6Q^\b+3U^\b+l^2\nabla^\ast_{\!\n}Y^{\n\b}
    \\
      \langle\; g_{\a\b}:\;\rangle & -i\planck^2\cdot j^\c &
      4l^2\cdot \nabla^\ast_{\!\n}Y^{\n\c}
    \\
      \langle\; 1/6 \cdot\eta_{\c\b\a\d}:\;\rangle
      &-\frac{1}{4}\planck^2\cdot j^{\SSst 5}{}_{\!\d} &
      -\frac{1}{12}V_\d
    \end{array}
  \] }
\hfill
\parbox{1cm}{\baq\eaq}\\
One can easily derive $-Q^\a=U^\a=-i\planck^2 / 4 \cdot j^\a$. Inserting
this and the last two equations of (3.5) into (3.4) we obtain
$\Upsilon_{\a\b\c}=0$. Since the Levi--Civita connection
fulfills $\chr{a}\m{b}+\chr{b}\m{a}=0$, where
$\chr{a}\m{b}=e_{\aa\a}\,e_\bb{}^{\!\b}\Chr\a\m\b+e_{\aa\c}\partial_\m
e_\bb{}^{\!\c}$, it follows
$\Con\aa\m\aa=\Sigma^\aa{}_{\!\m\aa}=Q_\m+4S_\m+U_\m
=4S_\m$ and thus $Y_{\m\n}=4S_{\m\n}:=4(\partial_\m S_\n-\partial_\n S_\m)$.
These results amount to
\baq
    \Con\a\m\b
  &=&
    {\widehat{\Gamma}}^\a{}_{\!\m\b}+\d^\a{}_{\!\b}\cdot S_\m
    \;,\,\mbox{ where}
  \\
    {\widehat{\Gamma}}^\a{}_{\!\m\b}
  &:=&
    \Chr\a\m\b + {\Tst\frac{1}{4}}\planck^2
    \left( i\cdot j^\a g_{\m\b}-i\cdot\d^\a{}_{\!\m}j_\b-
           \eta^\a{}_{\!\m\b\d} j^{{\SSst 5}\,\d}    \right)
\eaq
and\hfill
\parbox{10cm}{
  \[
    16i l^2 / \planck^2\,\nabla^\ast_{\!\n}S^{\n\c} = j^\c \;\;.
  \] }
\hfill\parbox{1cm}{\baq\eaq}\\
The last equation implies the current conservation
\beq
  \nabla^\ast_{\!\c}j^\c = 16i l^2 / \planck^2\,
                           \nabla^\ast_{\!\c}\nabla^\ast_{\!\n}S^{\n\c}
                         = 0 \;\;.
\eeq
So far we have not used the complex extension of the connection explicitly.
But now from (3.7) and (3.8) we see that parts of the connection (3.6) must
be complex valued. In other words, these equations can not be solved using a
real connection only. This is exactly the reason why we have chosen a complex
rather than a real linear
connection as our field variable. As shown in Appendix A the connection (3.6)
is a sum of the complex metric connection ${\widehat{\Gamma}}^\a{}_{\!\m\b}$
on the complex Lorentz bundle
\clplusm\ and the \uone--connection $S_\m$ on the trivial \uone--bundle
$M\times\uone$, glued together by canonical bundle mappings. If $S_\m$
is a \uone--connection, it must be purely imaginary, see (3.15). This can
not be deduced from (3.8) alone as it
contains only $\nabla^\ast_{\!\n}\mbox{Re}(S^{\n\c})=0$, but not
$\mbox{Re}(S_\m)=0$ itself. According to Appendix A the corresponding
\uone--gauge transformation is given by
\beq
  e_\aa{}^{\!\m}\mapsto e_\aa{}^{\!\m}\;,\quad
  {\Tst\frac{1}{4}}\Con\aa\m\aa (=S_\m)\mapsto
  {\Tst\frac{1}{4}}\Con\aa\m\aa+\partial_\m \l \;,\quad
  \psi \mapsto \exp(\l)\psi \;\;.
\eeq
\\[3mm]
{\bf (b)} \hspace{2mm} The Lagrangian (2.5) immediately yields
$0=\d{\cal L} / \d\PSI =\partial{\cal L} / \partial\PSI$ or, equivalently,
$i\c^\m\nabla_{\!\m}\psi-mc / \hbar\,\psi=0$. With (3.6)
this can be converted into
\beq
   i\c^\m(\nabla^\ast_{\!\m}-S_\m)\psi - \frac{mc}{\hbar}\psi
   +{\Tst\frac{3}{8}}\planck^2 (j_\m+j^{\SSst 5}{}_{\!\m}\c^5)\c^\m\psi
  =0 \;\;,
\eeq
where
\beq
  \nabla^\ast_{\!\m}\psi:=\partial_\m\psi
                         -{\Tst\frac{1}{4}}\chr{a}\m{b}\c^\bb\c^\aa\psi
\eeq
is the covariant spinor differentiation with respect to the Levi--Civita
connection. The spinor equation for $\PSI$ is more difficult to compute
\cite{12}. The result is
\beq
   i(\nabla^\ast_{\!\m}+S_\m)\PSI\cdot\c^\m + \frac{mc}{\hbar}\PSI
   -{\Tst\frac{3}{8}}\planck^2 \PSI (j_\m+j^{\SSst 5}{}_{\!\m}\c^5)\c^\m
  =0
\eeq
with $\nabla^\ast_{\!\m}\PSI=\overline{\nabla^\ast_{\!\m}\psi}$. The
nonlinear terms in (3.11) and (3.13) vanish due to the
identity
\beq
  (j_\m+j^{\SSst 5}{}_{\!\m}\c^5)\c^\m\psi=0,
\eeq
which can be derived by straightforward but cumbersome
computations, recalling $j_\m=\PSI\c_\m\psi$ and
$j^{\SSst 5}{}_{\!\m}=\PSI\c^{\SSst 5}\c_\m\psi$ and
using e.\ g.\ the chiral representation and the properties of
the Pauli matrices within the $\c$--matrices.

Since (3.13) is the spinor equation of the adjoint spinor $\PSI$, it must
agree with the adjoint of the first equation (3.11). This implies that
$S_\m$ is purely imaginary,
\beq
  \mbox{Re}(S_\m)=0 \;\;.
\eeq
\\[3mm]
{\bf (c)} \hspace{2mm} The Lagrangian (2.5) contains no derivatives of
$e_\aa{}^{\!\m}$ and therefore we get
\baq
    0
  &\!=\!&
    \frac{\d{\cal L}}{\d e_\cc{}^{\!\a}}e_{\cc\b}
    = \frac{\partial{\cal L}}{\partial e_\cc{}^{\!\a}}e_{\cc\b}
   \,=\,\left[-{\cal L}_m\,g_{\a\b}+e\,i\hbar
c\PSI\c_\a\nabla_{\!\b}\psi\right]
  \;\;\nonumber
  \\
  & &
    -\,{\Tst\frac{1}{2k}}\left[-eR\,g_{\a\b}+eR_\a{}^\m{}_{\!\b\m}
                               +eR^\m{}_{\!\a\m\b}                \right]
    +\,{\Tst\frac{1}{4k}}l^2\left[-eY_{\m\n}Y^{\m\n}g_{\a\b}
                                  +4eY_{\m\a}Y^\m{}_{\!\b}  \right] \;.\qquad
\eaq
Using (3.3), (3.6) to (3.8) and (3.11) to (3.14) this can be expressed as
\baq
  & & \qquad\qquad\qquad T^G_{\a\b}=T^m_{\a\b}+T^S_{\a\b} \;\;,
\\
    T^G_{\a\b}
  &\!:=\!&
    {\Tst\frac{1}{k}}
    \left( R^\ast_{\a\b}-{\Tst\frac{1}{2}}R^\ast g_{\a\b}\right) \;;
\\
    T^m_{\a\b}
  &\!:=\!&
    {\Tst\frac{i\hbar c}{2}}
    \left[ \PSI\c_\a(\nabla^\ast_{\!\b}-S_\b)\psi
          -(\nabla^\ast_{\!\b}+S_\b)\PSI\cdot\c_\a\psi
          +{\Tst\frac{1}{2}}\nabla^{\ast\c}(\PSI\c_{[\a}\c_\b\c_{\c]}\psi)
    \right] \;;
    \qquad
\\
    T^S_{\a\b}
  &\!:=\!&
    {\Tst\frac{16}{k}}l^2
    \left[
S_{\a\c}S_\b{}^\c-{\Tst\frac{1}{4}}S_{\m\n}S^{\m\n}g_{\a\b}\right]\;.
\eaq
In (3.18) $R^\ast_{\a\b}$ and $R^\ast$ denote the Ricci--tensor and --scalar
for the Levi--Civita connection. Since $T^G_{\a\b}$ and $T^S_{\a\b}$ are
symmetric in $\a$ and $\b$, (3.17) transfers this property also upon
$T^m_{\a\b}$. Indeed, a lengthy calculation \cite{12} gives
\beq
  T^m_{\a\b} =
  {\Tst\frac{i\hbar c}{4}}
  \left[ \PSI\c_\a(\nabla^\ast_{\!\b}-S_\b)\psi
        -(\nabla^\ast_{\!\b}+S_\b)\PSI\cdot\c_\a\psi
        +(\a \leftrightarrow \b) \right] \;\;.
\eeq
Since $T^G_{\a\b}$ is proportional to the Einstein--tensor, we obtain
\beq
  0 = \nabla^\ast_{\!\a}(T^G){}^{\a\b}
    = \nabla^\ast_{\!\a}(T^m){}^{\a\b}+\nabla^\ast_{\!\a}(T^S){}^{\a\b}\;\;.
\eeq
\setcounter{equation}{0}

%%%%%%%%%%%%%%%%%%%%%%%%%%%%%%%%%%%%%%%%%%%%%%%%%%%%%%%%%%%%%%%%%%%%%%%%%%%
%
                   \section{Physical interpretation}
%
%%%%%%%%%%%%%%%%%%%%%%%%%%%%%%%%%%%%%%%%%%%%%%%%%%%%%%%%%%%%%%%%%%%%%%%%%%%

We now summarize those field equations which will be discussed in the
following\\
\parbox{13mm}{}\hfill
\parbox{11cm}{
  \[
    16i l^2 / \planck^2\,\nabla^\ast_{\!\n}S^{\n\c} = j^\c
  \] }
\hfill\parbox{13mm}{(3.8')}\\[-2mm]
\parbox{13mm}{}\hfill
\parbox{11cm}{
  \[
    e_\aa{}^{\!\m}\mapsto e_\aa{}^{\!\m}\;,\quad
    {\Tst\frac{1}{4}}\Con\aa\m\aa (=S_\m)\mapsto
    {\Tst\frac{1}{4}}\Con\aa\m\aa+\partial_\m \l \;,\quad
    \psi \mapsto \exp(\l)\psi \;\;.
  \] }
\hfill\parbox{13mm}{(3.10')}\\[-2mm]
\parbox{13mm}{}\hfill
\parbox{11cm}{
  \[
    i\c^\m(\nabla^\ast_{\!\m}-S_\m)\psi - \frac{mc}{\hbar}\psi = 0
  \] }
\hfill\parbox{13mm}{(3.11')}\\[-2mm]
\parbox{13mm}{}\hfill
\parbox{11cm}{
  \[
    i(\nabla^\ast_{\!\m}+S_\m)\PSI\cdot\c^\m + \frac{mc}{\hbar}\PSI = 0
  \] }
\hfill\parbox{13mm}{(3.13')}\\[-2mm]
\parbox{13mm}{}\hfill
\parbox{11cm}{
  \[
    T^G_{\a\b}=T^m_{\a\b}+T^S_{\a\b}\;\;.
  \] }
\hfill\parbox{13mm}{(3.17')}

These field equations
exhibit precisely the well-known structures of the Einstein--Maxwell theory,
provided that $S_\m$ is identified with the electromagnetic potential $A_\m$
\beq
  S_\m = \frac{iq}{\hbar c} A_\m \;\;,
\eeq
where $q$ is the (positive) elementary charge.
In this case, (3.8') is simply the inhomogeneous Maxwell equation,
see (4.2), whereas (3.10') describes the
electromagnetic \uone-gauge transformation of a {\em negatively} charged
particle, which we identify with electron. The wave equation (3.11') becomes
the corresponding charged spinor equation in a curved space--time, (3.13')
being its adjoint. Note that $\nabla^\ast_{\!\m}-S_\m$ in (3.11')
is the \uone--gauge covariant spinor derivative.
Finally, (3.17') gives the energy--momentum
equation involving the energy--momentum tensors of gravity (3.18), charged
spinor particle (3.19) or (3.21), and the electromagnetic field (3.20).

In order to fix the length scale $l$ we insert (4.1) into (3.8') and
compare it with the usual Maxwell equation
\baq
    j^\c
  \stackrel{(4.1)}{=}
    16i l^2 / \planck^2\,\frac{iq}{\hbar c}\,\nabla^\ast_{\!\n}F^{\n\c}
  \stackrel{!}{=}
    \frac{1}{-q}\,\nabla^\ast_{\!\n}F^{\n\c}
  & \Leftrightarrow &
\nonumber\\
    l^2=\frac{1}{64\pi}\planck^2\frac{\hbar c}{q^2 / 4\pi}
       =\frac{1}{64\pi\,\a}\,\planck^2
  & \Rightarrow &
    l \approx 0.83 \,\planck \;\;,
\eaq
where $\a$ is the fine structure constant and we have employed
Heaviside--Lorentz units.
As expected in Section 2 the value of $l$ is of the same magnitude as the
Planck length, which indicates the close relation of electromagnetism to
space--time geometry and to gravity. If we had taken $l:=\planck$ in (2.5),
we would have obtained $\a=1 / 64\pi$ and $q\approx 1.32\cdot 10^{-19}
\mbox{Coulomb}$.
Renormalization procedures could perhaps improve (4.2) towards $l=\planck$.
When (4.1) and (4.2) are taken into account one can easily show that the
above mentioned
field equations of section 3 are exactly the equations of Einstein--Maxwell
theory with an electron. Moreover, with these results the Lagrangian (2.5)
can be rewritten as
\beq
  {\cal L} =
  e\cdot \hbar c
  \left[i\PSI\c^\m(\nabla^\ast_{\!\m}-\frac{iq}{\hbar c}A_\m)\psi
        -\frac{mc}{\hbar}\PSI\psi                              \right]
 -\frac{e}{2k}\,R^\ast
 -\frac{e}{4}\,F_{\m\n}F^{\m\n} \;\;,
\eeq
which is the usual Einstein--Maxwell Lagrangian.

Notwithstanding these agreements of our theory with the usual one there are
important differences, which will now be discussed. From
Appendix A we know that the ``gravitational'' metric connection
${\widehat{\Gamma}}^\a{}_{\!\m\b}$ (3.6) and the \uone-potential
$S_\m$ emerge out of a single connection by symmetry breaking.
In accordance with this geometrical background we regard $S_\m$ as the
electromagnetic vector potential rather than $A_\m$ itself. Therefore, we
describe the electromagnetic interaction through the field equations in
Section 3 together with the definite $l$ (4.2) only, thereby completely
disregarding (4.1). The problem of this geometrisation procedure,
implying the ``melting'' of $q$ and $A_\m$ into the single expression $S_\m$,
is, how to incorporate particles with charges different from $-q$. To solve
this problem we look closely at the spinor derivative (2.4). Assuming for a
moment that $\Con\aa\m\bb$ is metric, $\con\aa\m\bb=-\con\bb\m\aa$, we can
write (2.4) in the following three ways\\
\begin{minipage}{13.5cm}
  \[\renewcommand{\arraystretch}{2}
    \arraycolsep0.2mm
    \begin{array}{ @{\quad}c@{\quad}r@{\:}c@{\:}l@{\; =\;}l }
        \mbox{(+)}
      & \nabla_\m\psi
      & =
      & \partial_\m\psi-\frac{1}{4}\con\aa\m\bb\c^\bb\c^\aa \psi
      & \left( \partial_\m-\frac{1}{8}\con\aa\m\bb
               \left[\c^\bb , \c^\aa\right]
              -\frac{1}{4}\Con\aa\m\aa  \right) \psi
      \\
        \mbox{($-$)}
      &
      & =
      & \partial_\m\psi+\frac{1}{4}\con\aa\m\b\c^\aa\c^\bb \psi
      & \left( \partial_\m-\frac{1}{8}\con\aa\m\bb
               \left[\c^\bb , \c^\aa\right]
              +\frac{1}{4}\Con\aa\m\aa  \right) \psi
      \\
        \mbox{(0)}
      &
      &
      &
      & \left( \partial_\m-\frac{1}{8}\con\aa\m\bb
               \left[\c^\bb , \c^\aa\right] \right) \psi
        \;\;\mbox{.\rule[-8mm]{0mm}{10mm}}
    \end{array}
  \]
\end{minipage}
\hfill
\parbox{1cm}{\baq\eaq}\\
Turning back to our complex connection the first case ($+$) is exactly (2.4)
and describes a negatively charged particle, whereas ($-$) and (0)
correspond to positively charged and neutral particles with \uone--gauge
transformations $\psi\mapsto\exp(-\l)\psi$ and $\psi\mapsto\psi$ under (3.10),
respectively.

Consider now a many particle system. We distinguish the particles
with index $z$ and classify their charges according to (4.4) by the symbol
$\varepsilon=\varepsilon(z)$ taking three values
\beq
  \varepsilon=\varepsilon(z)=+,-,0.
\eeq
The Lagrangian density of this many particle system reads
\baq
    \widetilde{\cal L}
  &=&
    \sumd e\hbar c
    \left[ i\PSId\c^\m( \partial_\m
                           -{\Tst\frac{1}{8}}\con\aa\m\bb[\c^\bb,\c^\aa]
                           -\varepsilon(z){\Tst\frac{1}{4}}\Con\aa\m\aa ) \psid
          -\frac{m_z c}{\hbar}\PSId\psid
    \right] \nonumber
  \\
  & &
    -\frac{e}{2k}\,R+\frac{e}{4k}l^2\,Y_{\m\n}Y^{\m\n}
    \;\;\mbox{.}
\eaq
The field equations can be solved in the same manner as in section 3
yielding
\baq
    \Con\a\m\b
  &=&
    {\widetilde{\Gamma}}^\a{}_{\!\m\b}+\d^\a{}_{\!\b}\cdot S_\m
    \;,\,\mbox{ where}
  \\
    {\widetilde{\Gamma}}^\a{}_{\!\m\b}
  &:=&
    \Chr\a\m\b + {\Tst\frac{1}{4}}\planck^2
    \sumd\left( i\cdot j_z^\a g_{\m\b}-i\cdot\d^\a{}_{\!\m} j_{z\,\b}
               -\eta^\a{}_{\!\m\b\d} j_z^{{\SSst 5}\,\d}    \right)
\eaq
and\hfill
\parbox{10cm}{
  \[
    16i\,l^2 / \planck^2\,\nabla^\ast_{\!\m}S^{\m\n}
  = \sum^{}_{\varepsilon(z)=+}\currentd^\n\,
   -\sum^{}_{\varepsilon(z)=-}\currentd^\n \;\;,
  \] }
\hfill\parbox{1cm}{\baq\eaq}\\
where we have defined $\currentd^\a:=\PSId\c^\a\psid$ and $\currentd^{{\SSst 5}
\,\d}:=\PSId\c^5\c^\d\psid$. The energy--momentum equation (3.17) now becomes
\beq
  T^G_{\a\b}=\tilde{T}^m_{\a\b}+T^S_{\a\b}+W_{\a\b}
\eeq
with\hfill\raisebox{-3.5mm}{\parbox{13cm}{
  \baqn
    \tilde{T}^m_{\a\b}
  &=&
    \sumd\,e{\Tst\frac{i\hbar c}{4}}
    \big[ \PSId\c_\a(\nabla^\ast{}_{\!\b}-\varepsilon(z) S_\b)\psid
         -(\nabla^\ast{}_{\!\b}+\varepsilon(z) S_\b)\PSId\cdot\c^\a \psid
  \nonumber\\
  & &
    +(\a\leftrightarrow\b)\big]  \;\;,
  \eaqn }}
\hfill\parbox{1cm}{\baq\eaq}\\
and\hfill\parbox{13cm}{
  \baqn
    W_{\a\b}
  &=&
    e\,{\Tst\frac{3}{8k}}\planck^4
      \sum^{}_{z\neq z'}
      \left( j_{z\,\m}\,j_{z'}^\mu
            +j_{z \m}^{\SSst 5}\,j_{z'}^{{\SSst 5}\,\mu}
      \right) g_{\a\b} \;\;.\qquad\qquad\qquad\qquad
  \eaqn }
\hfill\parbox{1cm}{\baq\eaq}\\
The spinor equation (3.11) acquires a new term corresponding to (4.12)
\beq
    i\c^\m(\nabla^\ast_{\!\mu}-\varepsilon(z) S_\mu)\psid
   -\frac{m_z c}{\hbar}\psid
   +{\Tst\frac{3}{8}}\planck^2 \sum^{}_{z'\neq z}
    \left(j_{z'}^\m+j_{z'}^{{\SSst 5}\,\m}\c^5
    \right)\,\c_\m\psid
  = 0
  \;\;.
\eeq
We recognize that (4.9) is the correct inhomogeneous Maxwell equation of
the many particle system. Eqs.\ (4.13) are apart from the last
contribution the corresponding spinor equations for differently charged
particles. Thus, we could treat in a natural way the charges $\pm q$ and 0.

In (4.10) and (4.13) we can observe a spinor--spinor contact interaction
between distinct particles, to which both vector and axial currents
contribute. The absence of self-interactions among the spinors and also the
vanishing of the cubic terms in (3.11) and (3.13) are due to (3.14) and have
their origin in our special choice of ${\cal L}_m$ in (2.5), where we have
omitted the covariant differentiation of $\PSI$. Usually, the matter
Lagrangian is required to be real, neccesitating the inclusion of both
covariant derivatives of $\psi$ and $\PSI$. Since in our case the other
Lagrangians ${\cal L}_G$ and ${\cal L}_Y$ in (2.5) were already complex,
there was no need to make ${\cal L}_m$ alone real valued via the
consideration of the adjoint spinor derivative.
In Ref.\ \cite{2} a real Lagrangian
containing also this covariant derivative of $\PSI$ led to the self-inter%
action in (1.2), which is induced only via the axial current $j^{{\SSst 5}
\m}$. In our opinion, self-interactions of Dirac particles are unlikely
because of the Fermi--Dirac spin-statistics they obey and should be avoided.
We remark that spinor--spinor interactions are far too weak to be observed
by laboratory experiments, but can influence cosmological and quantum
gravitational phenomena \cite{1}.

Note that the presence of the vector current $j^\m$ also changes the simple
ansatz $T_\m\sim A_\m$ for the torsion trace, since from (3.6) we get
\beq
  T_\m = \Con\a\m\a-\Con\a\a\m = 3S_\m-3U_\m
       = 3\frac{iq}{\hbar c}A_\m+{\Tst\frac{3}{4}}i\planck^2\,j^\m \;\;.
\eeq

In addition to these features there is one more aspect of our theory which
differs from the usual Einstein--Maxwell theory. A Dirac spinor produces a
complex contorsion $1/4 \planck^2 (i\cdot j^\a g_{\m\b}
-i\cdot\d^\a{}_{\!\m}j_\b-\eta^\a{}_{\!\m\b\d} j^{{\SSst 5}\,\d})$
in ${\widehat{\Gamma}}^\a{}_{\!\m\b}$ (3.6), which
carries out the parallel transports of (uncharged) tensors on $M$. Since the
contorsion is a tensor, it does not vanish even in a local inertial system,
which we define to be the special coordinate system around a space--time
point $p\in M$ with $g_{\m\n}(p)=\mbox{diag}(1,-1,-1,-1)$ and $\partial_\s
g_{\m\n}(p)=0$. Although the physical meaning of this phenomenon is not
yet clear, we remark that it is invisible to the laboratory experiments of
today due to the very small magnitude of $\planck^2$ in the contorsion.
Furthermore, ${\widehat{\Gamma}}^\a{}_{\!\m\b}$ is a metric connection,
${\widehat{\nabla}}_{\!\m}g_{\a\b}=0$, where ${\widehat{\nabla}}_{\!\m}$ is
the covariant derivative defined in terms of
${\widehat{\Gamma}}^\a{}_{\!\m\b}$. This guarantees the
invariance and conserves the real nature of physical measurements of
lengths, time intervals, rest masses and various scalar products of
particle momenta.

\setcounter{equation}{0}

%%%%%%%%%%%%%%%%%%%%%%%%%%%%%%%%%%%%%%%%%%%%%%%%%%%%%%%%%%%%%%%%%%%%%%%%%%%
%
                           \section{Summary}
%
%%%%%%%%%%%%%%%%%%%%%%%%%%%%%%%%%%%%%%%%%%%%%%%%%%%%%%%%%%%%%%%%%%%%%%%%%%%

We have used a complex linear connection and an extended spinor derivative to
unify the gravitational and electromagnetic interactions into the
space--time geometry. Contrary
to other attempts at unification we could also clarify the fibre geometric
background (Appendix A).
The field equations are derived from a variational principle and exhibit
precisely the structures of the Einstein--Maxwell theory with a negatively
charged spinor particle. However, the many particle system reveals a new
type of spinor--spinor contact interaction, which occurs only between
distinct particles, explaining why it was absent in the single
particle case. Furthermore, our theory differs significantly from the usual
theory in the geometrical understanding and its physical
consequences. From the fibre bundle structure of our unification it follows
that a metric connection and an electromagnetic vector potential emerge
from the single complex linear connection by symmetry breaking. Accordingly,
we interpret electromagnetism purely geometrically and use as the only
physical constant a characteristic length $l$ close to the Planck length.
This geometrisation scheme involves the ``melting'' of the charge $q$ and
$A_\m$ into one single expression and so rules out the consideration of
particles with arbitrary charge. However, spinors with charges $\pm q$ and 0
could be treated in the theory using natural extensions of the covariant spinor
differentiation. Finally, the parallel transports of uncharged tensors on
space--time are carried out by the resultant metric connection mentioned
before, which contains a complex contorsion. Physical measurements
remain unaffected by this contribution because the metric is
preserved under the parallel displacements.

In our opinion, this complex contorsion gives first hints to a deeper
understanding of space--time structure although it is
only a very small contribution and unobservable in most cases.\\[4mm]

{\bf Acknowledgements. }{ }I am indebted to Prof. M.\ Kretzschmar for
his encouragement and helpful discussions. I would like to thank
Prof.\ G.\ Mack for many indirect supports. It is also a pleasure to
thank Prof.\ N.\ Papadopoulos for valuable discussions.
I am grateful to Dr.\ M.\ Tung and to Dr.\ P.\ O.\ Roll for the
careful reading of the manuscript.

\setcounter{equation}{0}

%%%%%%%%%%%%%%%%%%%%%%%%%%%%%%%%%%%%%%%%%%%%%%%%%%%%%%%%%%%%%%%%%%%%%%%%%%%%
%%%%%%%%%%%%%%%%%%%%%%%%%%%%%%%%%%%%%%%%%%%%%%%%%%%%%%%%%%%%%%%%%%%%%%%%%%%%
\begin{appendix}

%%%%%%%%%%%%%%%%%%%%%%%%%%%%%%%%%%%%%%%%%%%%%%%%%%%%%%%%%%%%%%%%%%%%%%%%%%%%
%
                        \section{Bundle Geometry}
%
%%%%%%%%%%%%%%%%%%%%%%%%%%%%%%%%%%%%%%%%%%%%%%%%%%%%%%%%%%%%%%%%%%%%%%%%%%%%

We derive the correct geometrical definition of (2.4) and clarify the
bundle structure of our unification. In Subsection A.1 we establish the
general fibre bundle structure of the theory and especially introduce the
extended spin structures essential to the building of the spinor derivative
(2.4). In A.2 this fibre geometry is used to obtain a special spin
connection $\o_s$ from a given complex linear connection $\o$. Considering
local cross sections in A.3, we will recognize that $\o_s$ indeed results
in the correct spinor derivative (2.4) when written in local form. Here we
also observe the symmetry breaking aspect of the connection (3.6). Finally,
the properties of the \uone--gauge transformation are explained in A.4.

For the general theory of fiber
bundles see Refs.\ \cite{13,14}, for details concerning the spin geometry
we refer to Refs.\ \cite{15,16} and \cite{12}. In the following all
structures are
$C^\infty$ or analytic. We denote the Lie group homomorphism and its Lie
algebra homomorphism by the same letter.

                     \subsection{Fibre Bundle Structure}

Analogously to the real case the complex frame bundle $F_c(M)$ can be
reduced to a special complex
Lorentz bundle \clplusm, the structure group being the special complex
Lorentz group \clplus,
\[
  \clplus := \left\{\Lambda\in\mbox{Mat}(4,\!\complex) |
  \Lambda^T\eta\Lambda=\eta\,,\:\det\Lambda=1 \right\}\;\;,
\]
which is isomorphic to the special orthogonal group $\mbox{SO}(4,\!\complex)$.
The spin group (more precisely the component topologically connected with the
\One) corresponding to \clplus\
will be denoted by \cspin ($\cong\sltwoc\times\sltwoc$) and the
accompanying twofold covering homomorphism by $\xi_o:\cspin\rightarrow
\clplus$. Using a spin representation $\z:\cspin\rightarrow\glfourc$ we
define an enlarged spin group $\cspin\times\complex^\times$ and
the corresponding extended spin representation $\z^\times$
\beq
  \z^\times : \cspin\times\complex^\times \rightarrow \glfourc \;,\;\;
              (A,c) \mapsto \z(A)\cdot c^{-1}  \;\;,
\eeq
where $\complex^\times$ ($\cong\glonec$) is the multiplicative group $\complex
\backslash\{0\}$. The representation $c^{-1}$ was chosen in order
to obtain (2.4). Other possible choices $c^{+1}$ and $c^0$ correspond to $(-)$
and $(0)$ of (4.4), respectively.
We further define the homomorphism $\theta_o$
by
\beq
  \theta_o : \clplus\times\complex^\times \rightarrow \glfourc \;,\;\;
             (\Lambda ,c) \mapsto \Lambda\cdot c \;\;.
\eeq
One can easily show $\mbox{Ker}(\theta_o)=\{(\One,1),(-\One,-1)\}$ and that
every image $\Lambda c$ has exactly two  inverse images $(\Lambda,c)$
and $(-\Lambda,-c)$. Denote the image group of $\theta_o$ by
$G:=\theta_o(\clplus\times\complex^\times)$ and its canonical
inclusion in \glfourc\ by $j_o$. Obviously, the Lie algebra of \clplus\ is
given by $\complex\otimes\fff{l}$, where $\fff{l}$ is the Lie algebra of
\lplus\ (2.1). The homomorphism
$\theta_o$ induces an isomorphism of the Lie algebra
$\complex\otimes\fff{l}\,\times\,\complex$ of $\clplus\times\complex^\times$
onto the Lie algebra \fff{g} of $G$, given by $\fff{g}=\complex\otimes
\fff{l}\oplus\complex\cdot\One$. We can now write down the following
diagram of Lie group homomorphisms
\beq
    \glfourc
  \stackrel{\z^\times}{\longleftarrow}
    \cspin\times\complex^\times
  \stackrel{\xi_o\times id}{\longrightarrow}
    \clplus\times\complex^\times
  \stackrel{\theta_o}{\longrightarrow}
    G
  \stackrel{j_o}{\longrightarrow}
    \glfourc
\eeq
and construct an analogous diagram of principal bundle mappings
\beq
    S^\times(M)
  \stackrel{(\ast)}{\longleftarrow}
    (\cspin\times\complex^\times)(M)
  \stackrel{\xi\times id}{\longrightarrow}
    (\clplus\times\complex^\times)(M)
  \stackrel{\theta}{\longrightarrow}
    G(M)
  \stackrel{j}{\longrightarrow}
    F_c(M)
  \;.
\eeq
If $G_i(M)$, $i=1,2$, are $G_i$--principal bundles, then $(G_1\times G_2)(M)$
is the $G_1\times G_2$--principal bundle given by the restriction of $G_1(M)
\times G_2(M)$ to the diagonal $\Delta \subset M\times M$, where $\Delta$
is identified with $M$ itself \cite{13}.
\cxm\ is the trivial $\complex^\times$--bundle $M\times
\complex^\times$, and \cspinm\ is a \cspin--bundle with the corresponding
spin structure $\xi:\cspinm\rightarrow\clplusm$ satisfying $\xi(uA)=\xi(u)
\xi_o(A)$ for $u\in\cspinm$ and $A\in\cspin$. $S^\times(M)$ is the extended
spinor bundle defined to be the associated vector bundle $S^\times(M)=(
\cspin\times\complex^\times)(M)\,\times_{\z^\times}\,\complex^4$, ($\ast$)
denoting this building procedure. Note that an element $\phi$ of $S^\times(M)$
is an equivalence class \cite{13}
\beq
    \phi=[u,\phi_o]=[u\cdot (A,c)\,,\:\z^\times(A,c)^{-1}\cdot\phi_o]\;\;,
\eeq
where $u\in (\cspin\times\complex^\times)(M)$, $\phi_o\in\complex^4$ and $(A,c)
\in\cspin\times\complex^\times$. $G(M)$ is the $G$--principal bundle
consisting of elements $(c\cdot X^a)$, where $c\in\complex^\times$ and $(X^a)
\in\clplusm$. It is thus contained in $F_c(M)$, $j$ denoting the canonical
inclusion. Finally, $\theta$ is defined as follows
\beq
  \theta : (\clplus\times\complex^\times)(M) \rightarrow G(M) \;,\;\;
           ((X^a),c) \mapsto (c\cdot X^a) \;\;.
\eeq
We remark that the bundle mappings $\xi\times id$, $\theta$, and $j$ have
as their corresponding Lie group homomorphisms exactly $\xi_o\times id$,
$\theta_o$, and $j_o$.

                   \subsection{Spin Connection}

Having explained (A.4) we now construct the spinor derivative (2.4) from a
single complex linear connection $\o$ on $F_c(M)$. First define the vector
subspace \fff{m} in the Lie algebra $\fff{gl}(4,\!\complex)$ of \glfourc
\beq
  \fff{m}:=\left\{ A\in\fff{gl}(4,\!\complex) | A^T\eta-\eta A=0 \;
                  \mbox{and}\;\:\mbox{Tr}(A)=0 \right\}
  \;\;.
\eeq
Then one can easily verify the following (vector space) decomposition\\
\parbox{13.8cm}{
  \[\renewcommand{\arraystretch}{1.5}
    \arraycolsep0.3mm
    \begin{array}{ccccccc}
      \fff{gl}(4,\!\complex)\;\; &=& \complex\otimes\fff{l} &\oplus&
      \complex\cdot\One &\oplus& \fff{m} \\
      A &=\;& \frac{1}{2}(A-\eta A^T\eta) &\;+\;&
      \frac{1}{4}\mbox{Tr}A\!\cdot\!\One &\;+\;& \frac{1}{2}(A+\eta A^T\eta
      -\frac{1}{2}\mbox{Tr}A\!\cdot\!\One)  \;.
    \end{array}
  \] }
\hfill
\parbox{1cm}{\baq\eaq}\\
Now, $\fff{g}=\complex\otimes\fff{l}\oplus\complex\cdot\One$ is the Lie
algebra of $G$, and for every $\Lambda c \in G$ it follows $(\Lambda c)
\fff{m}(\Lambda c)^{-1} \subset \fff{m}$. In this case the \fff{g}--component
of $\o|_{G(M)}$ is a $G$--connection on $G(M)$ (see Ref.\ \cite{13}), which we
denote by $\o_G$. Using the Lie algebra isomorphism $\theta_o$ we can
convert the pull--back $\theta^\ast\o_G$ into a connection $\theta_o^{-1}
\theta^\ast\o_G$ on $(\clplus\times\complex^\times)(M)$. By the same token
$\o_s:=(\xi_o\times id)^{-1}(\xi\times id)^\ast(\theta_o^{-1}\theta^\ast\o_G)$
is a connection on $(\cspin\times\complex^\times)(M)$, which we call the
extended spin connection.

Using the canonical bundle projections $f_{cl}:(\clplus\times
\complex^\times)(M)\rightarrow\clplusm$ and $f_{c\times}:\rightarrow\cxm$
together with the concomitant group projections $f_{o\,cl}:\clplus\times
\complex^\times\rightarrow\clplus$ and $f_{o\,c\times}:\rightarrow
\complex^\times$ we get the connections $\o_{cl}:=f_{o\,cl}(\theta_o^{-1}
\theta^\ast\o_G)|_{\footnotesize\clplusm}$ and $\o_{c\times}:=f_{o\,c\times}
(\theta_o^{-1}\theta^\ast\o_G)|_{\footnotesize\cxm}$, which can be used to
express $\theta_o^{-1}\theta^\ast\o_G$ in accordance with the fiber product
structure of the underlying bundle
\beq
  \theta_o^{-1}\theta^\ast\o_G =
  f_{cl}^\ast\o_{cl}\oplus f_{c\times}^\ast\o_{c\times} \;\;.
\eeq
This leads to a corresponding decomposition of the extended spin connection,
where we omit the bundle projections for the sake of simplicity
\beq
  \o_s = \xi^{-1}_o\xi^\ast\o_{cl}\oplus \o_{c\times} \;\;.
\eeq

                    \subsection{Spinor Derivative}

This spin connection $\o_s$ defines a covariant differentiation on the
associated vector bundle $S^\times(M)$. We now show that (2.4) is exactly
this derivative. Let ${\cal U}\subset M$ be an open set and introduce cross
sections $\hat{\s}$ in $\cspinm|_{\cal U}$ and $\hat{1}$ in $\cxm|_{\cal U}$,
where $\hat{1}$ prescribes to each $p\in {\cal U}$ the value $\hat{1}(p):=
(p,1)\in\cxm$. $\xi(\hat{\s})$ is a cross section in \clplusm, that is, a
complex tetrad. Although we have used in our theory only real tetrads, the
following considerations remain valid even if $\xi(\hat{\s})$ is complex
valued. Let $\Con\aa\m\bb\,dx^\m:=\left(\xi\left(\hat{\s}\right){}^\ast
\o\right){}^\aa{}_{\!\bb}$ be the $\fff{gl}(4,\!\complex)$--components of the
pulled--back connection. Using (A.8) we obtain
\baq
    \left(\xi(\hat{\s}){}^{\!\ast}\o_{cl}\right){}^\aa{}_{\!\bb}
  &=&
    {\Tst\frac{1}{2}}(\Con\aa\m\bb-\Gamma_{\bb\m}{}^\aa) \qquad\mbox{and}
\\
    {\hat{1}}^\ast\o_{c\times}
  &=&
    {\Tst\frac{1}{4}}\Con\cc\m\cc \;\;.
\eaq
For the whole connection $\theta_o^{-1}\theta^\ast\o_G$ the decomposition
(A.9) implies
\beq
  \left((\xi(\hat{\s}),\hat{1}){}^{\!\ast}
  \theta_o^{-1}\theta^\ast\o_G\right){}^\aa{}_{\!\bb}
 =
  {\Tst\frac{1}{2}}(\Con\aa\m\bb-\Gamma_{\bb\m}{}^\aa)
  +{\Tst\frac{1}{4}}\Con\cc\m\cc\cdot\d^\aa{}_{\!\bb} \;\;.
\eeq
The solution (3.6) has exactly this structure and can thus be understood as
a sum of two connections (A.11) and (A.12) on two different bundles
according to (A.9).

A Dirac spinor $\psi$ is a cross section in $S^\times(M)$. Since
$(\hat{\s},\hat{1})$ is an element of $(\cspin\times\complex^\times)(M)$, we
can write $\psi$ as an equivalence class according to (A.5)
\beq
  \psi = \left[(\hat{\s},\hat{1})\,,\,\psi_{(\hat{\s},\hat{1})}\right] \;\;,
\eeq
where $\psi_{(\hat{\s},\hat{1})}$ is a $\complex^4$--valued function on
$\cal U$, usually denoted simply by the same letter $\psi$ and referred to
as the Dirac spinor itself. This convention was already used in (2.4). The
local trivialization of $S^\times(M)$ in (A.14) allows us to express the
covariant derivative of $\psi$ through \cite{14}
\beq
  \nabla_\m\psi =
  \left[(\hat{\s},\hat{1})\,,\,\partial_\m\psi_{(\hat{\s},\hat{1})}
        +\z^\times\left((\hat{\s},\hat{1}){}^{\!\ast}\o_s\right)
        \cdot\psi_{(\hat{\s},\hat{1})} \right] \;\;,
\eeq
where
\baq
    \z^\times\left((\hat{\s},\hat{1}){}^{\!\ast}\o_s\right)
  &\stackrel{(A.10)}{=}&
    \z\circ\xi^{-1}_o\xi(\hat{\s})^\ast\o_{cl} \,\oplus\,
    (-\One)\cdot\hat{1}{}^\ast\o_{c\times}
  \nonumber
  \\
  &=&
    {\Tst\frac{-1}{4}}\c^\bb\c_\aa\cdot
    {\Tst\frac{1}{2}}(\Con\aa\m\bb-\Gamma_{\bb\m}{}^\aa)\,-\,
    {\Tst\frac{1}{4}}\Con\cc\m\cc\cdot\One
  \nonumber
  \\
  &=&
    -{\Tst\frac{1}{4}}\Con\aa\m\bb\c^\bb\c_\aa \;\;.
\eaq
This shows the required agreement with (2.4). Note that we have used here
the (usual) explicit form of the Lie algebra homomorphism $\z\xi^{-1}_o$,
see \cite{15,12}. The $-\One$ in front of $\hat{1}{}^\ast\o_{c\times}$ is
due to the special choice $c^{-1}$ in the representation (A.1), whereas the
important factor $1/4$ comes from the decomposition (A.8).

                       \subsection{U(1)--Gauge Transformation}

If we change the section $\hat{1}$ to $\exp(\l)\cdot\hat{1}$, $\l$ being a
\complex-valued function, then from standard theories on gauge
transformations it follows
\[
  \left(\exp\left(\l\right)\cdot\hat{1}\right){}^{\!\ast}\o_{c\times} =
  {\Tst\frac{1}{4}}\Con\cc\m\cc+\partial_\m\l
\]
and\hfill\parbox{13cm}{
\[
  \psi = \left[(\hat{\s},\hat{1})\,,\,\psi_{(\hat{\s},\hat{1})}\right]
       \stackrel{(A.1)}{=}
         \left[(\hat{\s},\exp(\l)\cdot\hat{1})\,,\,
               \exp(\l)\cdot\psi_{(\hat{\s},\hat{1})}\right] \;\;.
\] }
\hfill\parbox{1cm}{\baq\eaq}\\
Since this $\complex^\times$--gauge transformation takes place on \cxm\
the tetrad $\xi(\hat{\s})$ on \clplusm\ remains unchanged. It is easy to
show that the adjoint spinor $\PSI$ transforms to $\exp(\overline{\l})
\cdot\PSI$, and that in (2.5) only the matter Lagrangian ${\cal L}_m$ is
affected by the change and transforms to $\exp(\l+\overline{\l})\cdot
{\cal L}_m$. The invariance of ${\cal L}_m$ thus implies $\l+\overline{\l}=
0$ or $\exp(\l)\in\uone$. We must therefore replace \cxm\ by its reduced
bundle $\uone(M):=M\times\uone$.

\setcounter{equation}{0}

%%%%%%%%%%%%%%%%%%%%%%%%%%%%%%%%%%%%%%%%%%%%%%%%%%%%%%%%%%%%%%%%%%%%%%%%%%%%%
%
                    \section{4-Vector Decomposition}
%
%%%%%%%%%%%%%%%%%%%%%%%%%%%%%%%%%%%%%%%%%%%%%%%%%%%%%%%%%%%%%%%%%%%%%%%%%%%%%

Given a third rank tensor $\Sigma_{\a\b\c}$ define $\Upsilon_{\a\b\c}$, $Q_
\a$, $S_\b$, $U_\c$, and $V_\d$ as follows:\\
\parbox{1cm}{ }
\hfill
\parbox{12cm}{
  \[\renewcommand{\arraystretch}{1.8}
    \arraycolsep0.2mm
    \begin{array}{rc@{\Big(\,}cccccc@{\,\Big)\,}l}
        \Sigma_{\a\b\c}=:\,\frac{1}{18}{\Dst\Big[}
      & & 5 & \Sigma_\a{}^\e{}_{\!\e}
        & - & \Sigma^\e{}_{\!\a\e}
        & - & \Sigma^\e{}_{\!\e\a}
            & g_{\b\c}
      \\
      &+& - & \Sigma_\b{}^\e{}_{\!\e}
        & +5& \Sigma^\e{}_{\!\b\e}
        & - & \Sigma^\e{}_{\!\e\b}
            & g_{\a\c}
      \\
      &+& - & \Sigma_\c{}^\e{}_{\!\e}
        & - & \Sigma^\e{}_{\!\c\e}
        & +5& \Sigma^\e{}_{\!\e\c}
            & g_{\a\b}
        {\Dst \,\Big] + \Sigma_{[\a\b\c]}
                    + \Upsilon_{\a\b\c} }
    \end{array}
  \] }
\hfill
\parbox{1cm}{\baq\eaq}\\[-7mm]
\baq
  \qquad\!\! & =: & \quad
  Q_\a\,g_{\b\c}\;+\;S_\b\,g_{\a\c}\;+\;U_\c\,g_{\a\b}\;-\;
  {\Tst\frac{1}{12}}\eta_{\a\b\c\d}V^\d\;+\;\Upsilon_{\a\b\c} \;\;.
\eaq
For $V^\d$ we have then $V_\d=2\eta_{\a\b\c\d}\cdot\Sigma^{\a\b\c}$. From
(B.1) we conclude\\[-2mm]
\parbox{1cm}{ }
\hfill
\parbox{12cm}{
  \[\renewcommand{\arraystretch}{1.8}
    \arraycolsep0.2mm
    \begin{array}{rcccccccl}
        \Sigma_\a{}^\e{}_{\!\e}=\,\frac{1}{18}{\Dst\Big[\,}
      & & 20& \Sigma_\a{}^\e{}_{\!\e}
        & -4& \Sigma^\e{}_{\!\a\e}
        & -4& \Sigma^\e{}_{\!\e\a}
            &
      \\
      & & - & \Sigma_\a{}^\e{}_{\!\e}
        & +5& \Sigma^\e{}_{\!\a\e}
        & - & \Sigma^\e{}_{\!\e\a}
            &
      \\
      & & - & \Sigma_\a{}^\e{}_{\!\e}
        & - & \Sigma^\e{}_{\!\a\e}
        & +5& \Sigma^\e{}_{\!\e\a}
            &
        {\Dst \,\Big] + \Upsilon_\a{}^\e{}_{\!\e}
                    \quad\Leftrightarrow\quad
        \Upsilon_\a{}^\e{}_{\!\e}=0 \;\;.}
    \end{array}
  \] }
\hfill
\parbox{1cm}{\baq\eaq}\\
Similarly, $\Upsilon^\e{}_{\!\b\e}=\Upsilon^\e{}_{\!\e\c}=0$. Further observe:
\beq
    \Sigma_{[\a\b\c]}
  =
    \Sigma_{[\a\b\c]}
   +\Upsilon_{[\a\b\c]}
  \Leftrightarrow
    \Upsilon_{[\a\b\c]}
  =
    0
  \;\;\mbox{.}
\eeq
For the last term in (3.4) note $Y^{\n\c}=-Y^{\c\n}$ and
$\Chr\n\n\m=\partial_\m e$, so that
\beq
  \nabla^\ast_{\!\n}Y^{\n\c} = \partial_\n Y^{\n\c}+\Chr\n\n\m Y^{\m\c}
                              +\Chr\c\n\m Y^{\n\m}
                             = \frac{1}{e}\partial_\n (eY^{\n\c}) \;\;.
\eeq
For the axial current observe $1/6\,\eta_{\c\b\a\d}\c^\c\c^\b\c^\a
=i\c^5\c_\d$,
where $\c^5=i\c^0\c^1\c^2\c^3$, see e.\ g.\ \cite{2}.

\end{appendix}
%%%%%%%%%%%%%%%%%%%%%%%%%%%%%%%%%%%%%%%%%%%%%%%%%%%%%%%%%%%%%%%%%%%%%%%%%%%%%


\begin{thebibliography}{88}

\bibitem[1]{1} F.\ W.\ Hehl, P.\ von der Heyde, G.\ D.\  Kerlick
  and J.\ M.\ Nester, Rev.\ Mod.\ Phys.\ {\bf 48} (1976) 393.

\bibitem[2]{2} F.\ W.\ Hehl and B.\ K.\ Datta, J.\ Math.\
  Phys.\ {\bf 12} (1971) 1334.

\bibitem[3]{3} A.\ Einstein, {\it The Meaning of Relativity}, Appendix II of
  the third ed., Princeton University Press, Princeton, 1950.

\bibitem[4]{4} L.\ Infeld, Acta Phys.\ Pol.\ {\bf X} (1950) 284.

\bibitem[5]{5} J.\ Callaway, Phys.\ Rev.\ {\bf 92} (1953) 1567.

\bibitem[6]{6} K.\ Borchsenius, Gen.\ Rel.\ Grav.\ {\bf 7} (1976) 527.

\bibitem[7]{7} J.\ W.\ Moffat and D.\ H.\ Boal, Phys.\ Rev.\ {\bf D11} (1975)
  1375; J.\ W.\ Moffat, Phys.\ Rev.\ {\bf D15} (1977) 3520.

\bibitem[8]{8} R.\ J.\ McKellar, Phys.\ Rev.\ {\bf D20} (1979) 356.

\bibitem[9]{9} M.\ Ferraris and J.\ Kijowski, Lett.\ Math.\ Phys.\ {\bf 5}
  (1981) 127.

\bibitem[10]{10} A.\ Jakubiec and J.\ Kijowski, Lett.\  Math.\ Phys.\ {\bf 9}
  (1985) 1.

\bibitem[11]{11} A.\ Ashtekar, J.\ D.\ Romano and R.\ S.\ Tate, Phys.\ Rev.\
  {\bf D40} (1989) 2572.

\bibitem[12]{12} K.\ Horie, {\it Die Vereinheitlichung von Gravitation und
  Elektromagnetismus durch die Torsion der Raum--Zeit}, thesis
  submitted to Johannes Gutenberg Universit\"at Mainz, 1994.

\bibitem[13]{13} S.\ Kobayashi and K.\ Nomizu, {\it Foundations of
  Differential Geometry} Vol.\ I, John Wiley \& Sons, New York, 1963.

\bibitem[14]{14} M.\ Nakahara, {\it Geometry, Topology and
  Physics}, Adam Hilger, Bristol, 1990.

\bibitem[15]{15} H.\ B.\ Lawson and M.-L.\ Michelsohn, {\it
  Spin Geometry}, Princeton Univ.\ Press, Princeton, 1989.

\bibitem[16]{16} D.\ Bleecker, {\it Gauge Theory and Variational
  Principles}, Addison--Wesley, Massachusetts, 1981.

                        \end{thebibliography}
\end{document}